\documentclass[useAMS,usenatbib]{mn2e} %
\pdfoutput=1

\usepackage{amsmath,amssymb,graphicx}
\usepackage{bm,color}
\usepackage{epstopdf}

\graphicspath{{figs/}}

\bibpunct{(}{)}{;}{a}{}{,}

\newcommand{\dd}{{\mathrm d}}

\newcommand{\Omegam}{\Omega_{\rm m}}

\newcommand{\Omegab}{\Omega_{\rm b}}

\title
[Bayesian model comparison in cosmology with PMC]
{Bayesian model comparison in cosmology with Population Monte Carlo}

\author[Kilbinger et al.]
{
  \parbox[h]{\textwidth}{
    Martin Kilbinger$^{1,2}$\thanks{E-mail: kilbinger@iap.fr},
    Darren Wraith$^{3,1}$,
    Christian P.~Robert$^3$,
    Karim Benabed$^1$,
    Olivier Capp\'e$^4$,
    Jean-Fran\c{c}ois Cardoso$^{4,1}$,
    Gersende Fort$^4$,
    Simon Prunet$^1$,
    Fran\c{c}ois R.~Bouchet$^1$
}
  \vspace*{10pt} \\
  \hspace{-.1cm}$^1$ Institut d'Astrophysique de Paris, UMR 7095 CNRS \&
  Universit\'e Pierre et Marie Curie, 98 bis boulevard Arago, 75014 Paris, France\\
  \hspace{-.1cm}$^2$ Shanghai Key Lab for Astrophysics, Shanghai Normal University,
  Shanghai 200234, P.~R.~China\\
  \hspace{-.1cm}$^3$ CEREMADE, Universit\'e Paris Dauphine, 75016
  Paris, France\\
  \hspace{-.1cm}$^4$ LTCI, Telecom ParisTech and CNRS, 46, rue Barrault,
  75013 Paris, France
}

\begin{document}

\date{\today}

\pagerange{\pageref{firstpage}--\pageref{lastpage}} \pubyear{2009}

\maketitle

\label{firstpage}

\begin{abstract}

  We use Bayesian model selection techniques to test extensions of the
  standard flat $\Lambda$CDM paradigm. Dark-energy and curvature
  scenarios, and primordial perturbation models are considered. To
  that end, we calculate the Bayesian evidence in favour of each
  model using Population Monte Carlo (PMC), a new adaptive sampling
  technique which was recently applied in a cosmological context. In
  contrast to the case of other sampling-based inference techniques
  such as Markov chain Monte Carlo (MCMC), the Bayesian evidence is
  immediately available from the PMC sample used for parameter
  estimation without further computational effort, and it comes with
  an associated error evaluation. 
  Also, it provides an unbiased
  estimator of the evidence after any fixed number of iterations and
  it is naturally parallelizable, in contrast with MCMC and nested
  sampling methods. By comparison with analytical predictions for
  simulated data, we show that our results obtained with PMC are
  reliable and robust. The variability in the evidence evaluation and
  the stability for various cases are estimated both from simulations
  and from data. For the cases we consider, the log-evidence is
  calculated with a precision of better than $0.08$.

  Using a combined set of recent CMB, SNIa and BAO data, we find
  inconclusive evidence between flat $\Lambda$CDM and simple
  dark-energy models. A curved universe is moderately to strongly
  disfavoured with respect to a flat cosmology. Using physically
  well-motivated priors within the slow-roll approximation of
  inflation, we find a weak preference for a running spectral index. A
  Harrison-Zel'dovich spectrum is weakly disfavoured. With the
    current data, tensor modes are not detected; the large
  prior volume on the tensor-to-scalar ratio $r$ results in moderate
  evidence in favour of $r=0$.
\vspace*{2em}
\end{abstract}


\section{Introduction}
\label{sec:intro}

We have reached an era of precision cosmology, as impressive
constraints on cosmological parameters attest \cite[e.g.][among many
others]{WMAP5-Dunkley08}. Parameters of the standard $\Lambda$CDM
model are measured with uncertainties of a few percent. At the same
time, we have not made the transition to what
\citet{2002astro.ph..8037P} called `accurate cosmology'. This next and
qualitatively different step involves the scrutiny of the underlying
model rather than the ever more precise determination of model
parameters. This approach is particularly important in the field of
cosmology which relies on a considerable extrapolation of known
physics to large scales and high energies (in the early Universe),
and lacks physical understanding, e.g.~of the dark
sector. Subsequently, cosmology has spawned a multitude of
different models.

There are several statistical approaches for comparing competing
models. Most of them aim at a balance between the ability of a
model to fit observational data, and its complexity.  While
information theory approaches like Akaike's criterion
\citep[AIC,][]{akaike:1974}
provide explicit penalisations for the complexity of a model, Bayesian
analysis compares directly the posterior probabilities of models,
in favour of each of one of the given
models. In that sense, Bayesian analysis has often been argued to
propose an automated Occam's razor, see
e.g. \cite{berger:jefferys:1992}, or \cite{mackay:2002}. Following the
classical Bayesian approach \citep{jeffreys:1939}, the comparison
between models integrates out the parameters within each model and
automatically penalises larger parameter spaces.

From a practical viewpoint, especially for high-dimensional parameter
spaces, the calculation of the evidence is very challenging. While
fast approximations exist, such as the Bayesian Information Criterium
\citep[BIC, ][]{schwarz78}, Laplace
\cite[see e.g.][]{2007MNRAS.380.1029H}, or variational Bayes
\cite[e.g.][]{mackay:2002}, they can fail dramatically for posterior
distributions which are not well approximated by a multivariate
Gaussian.

Recently, a new adaptive importance sampling method called Population
Monte Carlo (PMC) was introduced
\citep{cappe:guillin:marin:robert:2003,cappe:douc:guillin:marin:robert:2007}
and successfully tested in a cosmological context
\citep[][hereafter WKB09]{WK09}. PMC has since been used for a
  range of applications in cosmology \citep{2009MNRAS.400..219B,MKS09,SHJKS09}.

WKB09 focused on parameter estimation with the PMC sampling
algorithm.  In this second paper, we use the PMC method to estimate
the Bayesian evidence and assess the accuracy and reliability of
  this estimate. We emphasize that the same set of sampled
  values used for parameter estimation can also be used to calculate
the Bayesian evidence. Thus with the PMC method, model selection
comes at the same computational cost as parameter estimation.

This paper is organised as follows:
We describe the basics of the PMC
algorithm in Sect.~\ref{sec:model-selection-PMC}. Section \ref{sec:sims} assesses the
performance and reliability of PMC to calculate the Bayesian evidence
using numerical simulations. In Sect.~\ref{sec:cosmo} we use PMC to
compare cosmological models in the context of dark energy and
primordial perturbations. Our findings are summarised in
Sect.~\ref{sec:discussion}.

\section{Bayesian model selection with Population Monte Carlo}
\label{sec:model-selection-PMC}

\subsection{Bayesian evidence and Bayes' factor}

The Bayesian evidence $E$ in favour of a model $\mathfrak{M}$ with
likelihood function $L$ and prior distribution $P$ is the average of
the likelihood function, weighted by the prior, over the parameter space
\begin{equation}
 E = \int L(x) P(x) \, \dd x = \int \pi(x) \, \dd x,
 \label{evidence}
\end{equation}
where $\pi(x) = L(x)P(x)$ is the unnormalised posterior. We denote the
normalised posterior density by $\bar \pi(x)$, the normalisation constant
being the evidence, $\bar \pi(x) = \pi(x)/E$.

The Bayes factor \citep{jeffreys:1939} is the ratio of the evidences
$E_1$ and $E_2$ for two
competing models $\mathfrak{M}_1$ and $\mathfrak{M}_2$, respectively,
\begin{equation}
  B_{12} = \frac{E_1}{E_2}.
  \label{B_12}
\end{equation}
If $B_{12}$ is larger (smaller) than unity, the data favour model
$\mathfrak{M}_1$ ($\mathfrak{M}_2$) over the alternative model.  To quantify
the `strength of evidence' contained in the data,
\citet{jeffreys:1961} introduced an empirical scale, see Table \ref{tab:jeffrey}.
This is only a rough guideline for decision making and of course the
proposed boundaries are mostly subjective.
For a comprehensive review of Bayesian model selection, we refer the
interested reader to \cite{chen:shao:ibrahim:2000} and \citet{2008ConPh..49...71T}.

\begin{center}\begin{table}
  \caption{Jeffrey's scale to quantify the `strength of evidence' for
    a corresponding range of the Bayes factor $B_{12}$ in (\ref{B_12}), assuming $B_{12} > 1$.}
  \label{tab:jeffrey}
  \begin{tabular}{lll} \hline\hline
    $\ln B_{12}$ & $B_{12}$ & Strength \\ \hline
    $< 1$        & $< 2.7$ & Inconclusive \\
    $1 \ldots 2.5$ & $2.7 \ldots 12$ & Weak \\
    $2.5 \ldots 5$ & $12 \ldots 150$ & Moderate \\
    $>5$           & $>150$ & Strong
  \end{tabular}
\end{table}\end{center}

\subsection{A note on priors}

The Bayesian evidence depends crucially on the prior distribution on
the parameters. Firstly, in the Bayesian framework, the prior is an
integral part of the model, since Bayesian inference automatically
yields the \emph{updated} results with respect to prior knowledge.
Secondly, the concept of predictability or complexity of a model makes
sense only in comparison with the model prior; this is the core of the
Lindley-Jeffrey paradox \citep{lindley57} which is illustrated nicely
in \cite{2007MNRAS.378...72T}.  In short, a model is penalised if it
requires fine-tuning of parameters, corresponding to a posterior
  distribution that is very concentrated in terms of the prior mass.
\citep{2008MNRAS.388.1314E}. As \cite{2008MNRAS.388.1314E} pointed
out, the lack of a physically well-motivated model and therefore the
choice of an ad-hoc prior will strongly decrease the usefulness of a
model selection analysis. Since fundamental physical understanding is
often lacking in cosmology, the application of the Bayesian evidence
might indeed be limited. However, we should not consider this as a
fallacy of Bayesian inference but rather take it as a motivation to
find well-defined physical models which can be compared in a sensible
way \citep{2008ConPh..49...71T}.

\subsection{Estimating {the Bayesian} evidence with importance sampling}
\label{sec:PMC}

We propose to estimate the evidence using importance sampling (IS).
IS provides a converging approximation of the integral
(\ref{evidence}) as follows. For a probability density function $q$
whose support includes that of the posterior $\pi$, we can transform
(\ref{evidence}) as
\begin{equation}
  E = \int \pi(x) \, \dd x = \int \frac{\pi(x)}{q(x)} \, q(x) \, \dd x.
\end{equation}
IS performs a Monte-Carlo integration of $E$ by drawing $N$
samples $x_1, \ldots x_N$ from the \emph{importance function} $q$
to approximate $E$ with the sample average
\begin{equation}
  \label{eq:ISEstimOfE}
  E \approx \frac 1 N \sum_{n=1}^N w_n; \;\;\; w_n = \frac{\pi(x_n)}{q(x_n)},
\end{equation}
where the $w_n$ are called (unnormalised) importance weights. For
  later use, we introduce their normalised counterparts, marked
with a bar,
\begin{equation}
  \bar{w}_n = \frac{w_n}{\sum_{m=1}^{N}{w_m}}.
  \label{wbar}
\end{equation}
Of course, the quality of the estimate \eqref{eq:ISEstimOfE} 
of $E$ depends on the choice of the importance function $q$.
The variance of \eqref{eq:ISEstimOfE} is
\begin{equation}
  \sigma_E^2 = \frac{E^2} N \, d^2(\bar\pi\|q),
  \label{sigmaE}
\end{equation}
with the so-called chi-square distance
\begin{equation}
  \label{eq:chi2}
  d^2(\bar\pi\|q) = \int \frac{\bar\pi^2(x)}{q(x)}\, \dd x - 1.
\end{equation}
Therefore, a suitable choice of $q$ is such that $d^2(\bar \pi \| q)$
is as small as possible. In practice, this means that for each
specific problem (i.e., each specific $\pi$), efficient importance
functions have to be found, which is a non-trivial task. Hence the
appeal of \emph{adaptive} IS methods which rely on numerical
optimization schemes that can automatically select a suitable $q$.  In
the next section we discuss an algorithm based on an alternative
measure of fit between $\bar{\pi}$ and $q$, for which closed
adaptation expressions can be devised for $q$ being a mixture of
Gaussian or Student-t distributions.

\subsection{Adaptive importance sampling}

Population Monte Carlo tackles the problem of the importance function
choice by an adaptive solution: The PMC algorithm produces a sequence
$q^t, \, t=1,\ldots,T$ of importance functions aimed at
progressively approximating the posterior $\pi$.

The quality of approximation is measured in terms of the
Kullback-Leibler divergence
\citep{kullblack:leibler:1951,Cover1991} from the posterior,
\begin{equation}
  K({\bar\pi}\|q^t) = \int \log\left(\frac{{\bar\pi(x)}}{q^t(x)}\right) {\bar\pi(x)} \, \dd x,
  \label{eqn:kdiv}
\end{equation}
rather than the chi-square distance \eqref{eq:chi2},
and the density $q^t$ can be adjusted incrementally to minimize this
divergence. Note that the optimisation algorithm is independent of the
normalisation of the posterior. Therefore, for all practical purposes,
the unnormalised posterior $\pi$ can be used in (\ref{eqn:kdiv}).

The importance function should be selected from a family of functions
which is sufficiently large to allow for a close match with $\bar \pi$ but
for which the minimization of \eqref{eqn:kdiv} is computationally
feasible. \cite{cappe:douc:guillin:marin:robert:2007} propose to use
mixture densities of the form
\begin{equation}
  q^t(x) = q(x;\alpha^t,\theta^t) = \sum_{d=1}^D \alpha^t_d \,
  \varphi(x;\theta^t_d),
  \label{eq:mixtureISdensity}
\end{equation}
where $\alpha^t = (\alpha_1^t, \ldots, \alpha_D^t)$ is a vector of adaptable
weights for the $D$ mixture components (with $\alpha_d^t > 0$ and
$\sum_{d=1}^D \alpha_d^t = 1$), and $\theta^t = (\theta_1^t, \ldots,
\theta_D^t$) is a vector of parameters which specify the components;
$\varphi$ is a parametrized probability density function, usually taken
to be multivariate Gaussian or Student-t.
This choice of the importance function is very flexible and allows to
approximate a wide range of posteriors.

The first sample in this adaptive scheme 
is produced by a regular importance sampling mechanism,
$x_1^{1},\ldots,x_N^{1}\sim q^1$, associated with importance weights
\begin{equation}
  {w}_n^1 = \frac{\pi(x_n^{1})}{q^1(x_n^{1})};
  \quad n=1,\dots,N,
\end{equation}
providing a first approximation to a sample from $\pi$.

At each stage $t$ of the iteration, the sample points and weights from
that iteration, $(x_1^{t}, w_1^{t}), \dots, (x_N^{t}, w_N^{t})$, are
used to update the importance function $q^t$ to the new one,
$q^{t+1}$.  During the next iteration, a new sample
$x_1^{t+1},\ldots,x_N^{t+1}$ is then drawn from the updated importance
function $q^{t+1}$.

The updating method is based on a variant of the
Expectation-Maximization algorithm
\citep[EM,][]{dempster:laird:rubin:1977}. New parameters
$\alpha^{t+1}, \theta^{t+1}$ are obtained by carrying out an IS
approximation of the update of the EM algorithm to obtain a reduction
in the Kullback-Leibler divergence (\ref{eqn:kdiv}).

As an illustration of a simple updating rule, the new weights of the
mixture components can be calculated according to
\begin{align}
 \alpha^{t+1}_{d} & = \sum_{n=1}^N\bar{w}_{n}^t \, 1_d(x_n^t).
\label{eq:update_alpha} 
\end{align}
Here, $1_d(x)$ denotes the $d^{\rm th}$-component indicator function
which is unity if $x$ has been drawn from component $d$, and zero
otherwise. The updated component weight $\alpha_d^{t+1}$ is thus the
sum of the normalized importance weights of the sample points drawn
from the $d^{\rm th}$ component, leading to a simple intuition of the
adaptation. Points sampled from a component which approximates the
posterior well (badly) have large (small) weights, and the component
will be up-weighted (down-weighted) in the update.  Note that in our
implementation of PMC we use improved and more robust versions of
this updating rule. Details of the algorithm as well as formulae and
their derivations for updating the mean and covariance in the case of
Gaussian and Student-t mixtures are given in
\cite{cappe:douc:guillin:marin:robert:2007} and WKB09.

\subsection{Diagnostics}

Although a formal stopping rule for the above described iterative
process does not exist, performance measures can be defined to serve
as guidelines. As PMC aims at minimizing the Kullback-Leibler
divergence $K$ (\ref{eqn:kdiv}) across iterations, one can stop the
process when subsequent importance functions do not yield a
significant decrease of $K$. We estimate
$\exp[-K(\bar\pi\|q^t)]$ by the \emph{perplexity}
\begin{equation}
  p = \exp(H_{\rm N}^t)/N,
  \label{perplexity}
\end{equation}
where
\begin{equation}
  H_{\rm N}^t = -\sum_{n=1}^N \bar{w}_{n}^t\log\bar{w}_{n}^t
  \label{shannon-entropy}
\end{equation}
is the Shannon entropy of the normalised weights. Values of $p$ close
to unity will therefore indicate good agreement between the importance
function and the posterior.

Another frequently used criterion for importance sampling is the
so-called \emph{effective sample size }(ESS),
\begin{equation}
  \operatorname{ESS}_{\rm N}^t = \left( \sum_{n=1}^{N}
  \left\{\bar{w}_n^t\right\}^2 \right)^{-1},
  \label{ess}
\end{equation}
with $1 \le \operatorname{ESS}_{\rm N}^t \le N$. The effective sample size can be 
interpreted as the number of sample points with non-zero weight
\citep{liu:chen:1995}.
Both measures (\ref{perplexity}, \ref{ess}) are
related, as an importance function which is close to the
posterior density will in general have both a high perplexity and a
relatively large number of points with non-zero weight, compared to an
ill-fitting importance function.

\subsection{The initial proposal}

The efficiency of the algorithm is dependent on the initial choice of the proposal.
A poor initial importance function, e.g.~a single-mode function in the
case of a multimodal posterior or a too narrow function with light
tails, may take a very long time to adapt or even miss important 
parts of the posterior. For importance sampling the choice of $q$ requires
both fat tails and a reasonable match between $q$ and the posterior
$\pi$ in regions of high density. Such an importance function can be
more easily constructed in the presence of a well-informed guess about
the parameters and possibly the shape of the posterior density. In our
application of PMC to cosmology (Sect.~\ref{sec:cosmo}) we will use
the Fisher matrix as an aid for the initial proposal.

\subsection{Summary}

PMC offers a fast and reliable way to estimate the Bayesian evidence,
see eq.~(\ref{eq:ISEstimOfE}), with an expression of the variance of
this estimator provided by (\ref{sigmaE}). Diagnostics of the
reliability of the sampling are at hand.  The evaluations of the
likelihood function can further be massively parallelized with
importance sampling, offering an enormous decrease of the required
wall-clock time to obtain the evidence. This feature is almost unique
to importance sampling techniques and thus not partaken by alternative
techniques such as nested sampling and MCMC. We stress again that the
calculation of the evidence comes at virtually no further computation
cost than parameter estimation using PMC.

\section{Simulations}
\label{sec:sims}

In this section we use simulated data and a toy model for the
posterior density to assess the performance of the PMC approach to
provide an accurate estimate of the evidence. We use a non-Gaussian 
posterior density, twisting a centred $d$-dimensional multivariate Gaussian
in the first two dimensions.
One can easily build a sample under
this twisted posterior distribution, using
\begin{equation}
  x^\prime = (x^\prime_1,x^\prime_2,\ldots,x^\prime_d) \sim \mathcal{N}_d(0,\Sigma),
\end{equation}
where $\Sigma=\textup{diag}(\sigma_1^2,1,\dots,1)$ is the covariance,
and transforming it to
\begin{equation}
  x = (x^\prime_1,x^\prime_2-\beta ({x^\prime_1}^2-
  \sigma_1^2),x^\prime_3,\dots,x^\prime_{d}).
  \label{eqn:banana}
\end{equation}
The twist parameter $\beta$ controls the degree of curvature.
This example was also used in WKB09 to assess the
performance of PMC for parameter estimation.
In the following, we will use $d=5$ and $\sigma_1^2 = 100$.

\subsection{Unknown twist $\beta$}

As a first benchmark, our interest is in estimating $\beta$ and in obtaining the 
evidence $E$ by integrating out $\beta$ from the unnormalised posterior distribution, i.e.
\begin{equation}
  E=\int \pi(\beta|x,\Sigma) \, \dd\beta.
\end{equation}
As $\beta$ is a scalar we can easily calculate (using a grid-point
or adaptive quadrature approach) the evidence by integrating over
this one-dimensional domain.

\begin{figure}
  \resizebox{\hsize}{!}{
    \includegraphics{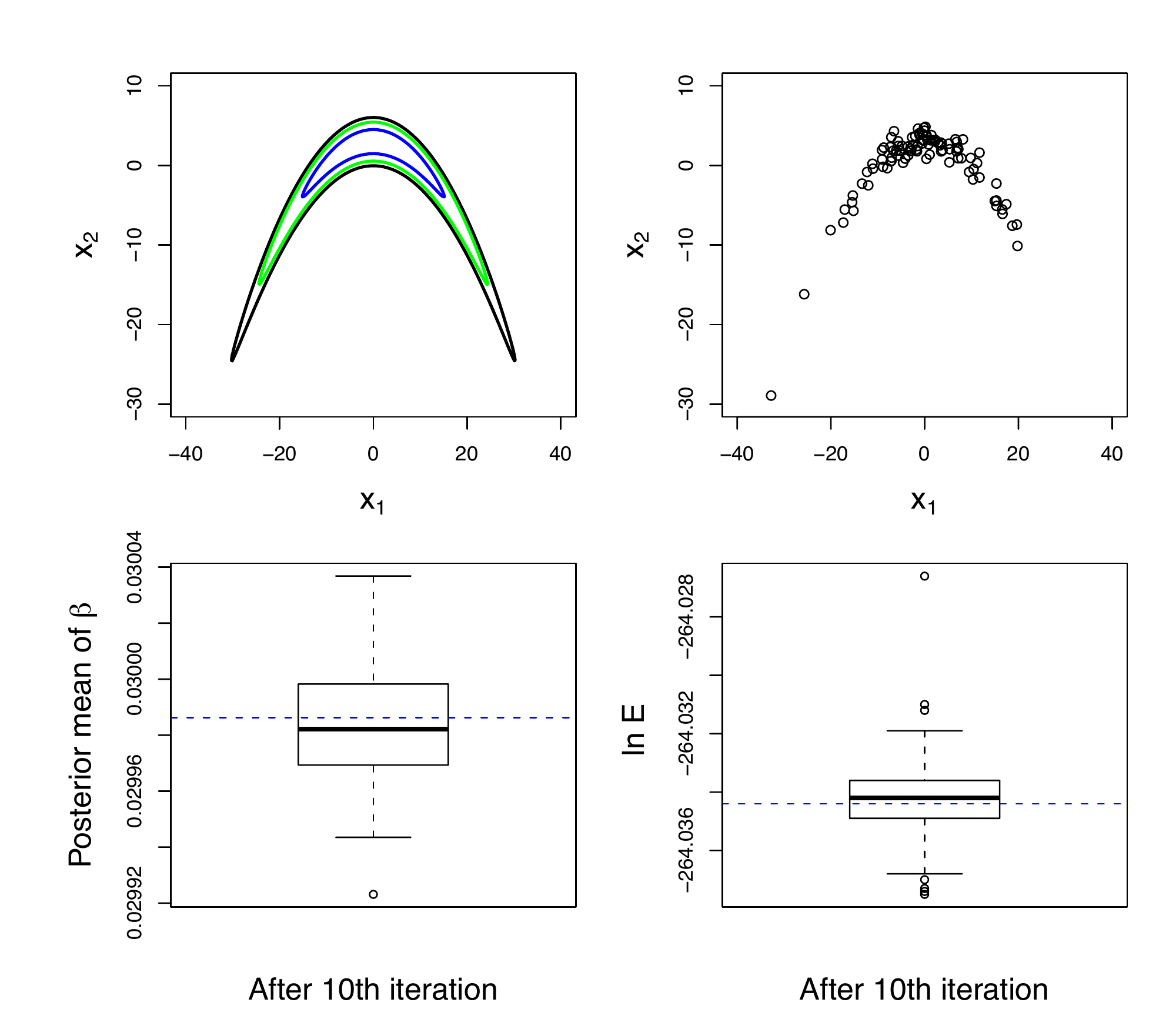}
  } 
  \caption{ {\em Top panels:} The posterior distribution
      (\ref{eqn:banana}) for $\beta=0.03$, with the true 68.3\%
      (blue), 95\% (green), and 99.7\% (black) density contours
      (\emph{left}) and $M=100$ simulated data points drawn from
      (\ref{eqn:banana}), both shown in the first two dimensions.
      {\em Bottom panels:} Estimates over 100 simulation runs after
      the 10$^{\rm th}$ iteration of $\beta$
      (\emph{left}), and the Bayesian evidence (\emph{right}).
      The distributions are
      shown as whisker plots: the thick horizontal line represents the
      median; the box shows the interquartile range (IQR), containing
      50\% of the points; the whiskers indicate the interval
      $1.5\times$IQR from either Q1 (lower) or Q3 (upper); points
      outside the interval $[Q1, Q3]$ (outliers) are represented as
      circles. Posterior means of $E$ and $\beta$ from the simulated
      data are indicated as dashed lines.}
  \label{fig:xsample}
\end{figure}

For this example we take a simulated data set $x$ with size $M=100$, and input 
twist $\beta=0.03$. Fig.~\ref{fig:xsample} shows the confidence contours of
the posterior density defined by eq.~(\ref{eqn:banana}) (top left
panel) and the simulated data points (top right) in the first two
dimensions. From eq.~(\ref{eqn:banana}), the likelihood is defined as
\begin{equation}
L(\beta)=\prod_{m=1}^M
\frac{1}{(2\pi)^{d/2}|\Sigma|^{1/2}}\text{exp}(-0.5\, x_m^{\rm T}\Sigma^{-1}x_m),
\end{equation}
where each $x_m = (x_{m1},x_{m2}+\beta (x_{m1}^2-
\sigma_1^2),x_{m3},\dots,x_{md})$ is a transformed sample point
according to \eqref{eqn:banana}. Note that the Jacobian of the
transformation is unity. The prior for $\beta$ is uniform
on the unit interval, $P(\beta)=U(0,1)$.

The initial importance function $q^1$ was chosen to be a mixture of
three Gaussian distributions with means randomly displaced around zero
and variance $0.5$, allowing for a somewhat vague initial coverage of
the parameter space.  For the results to follow the number of points
at each PMC iteration was $N=1000$, with a number of PMC iterations
equal to $T=10$.  To assess the variability and distribution of the
results we repeated this process 100 times.  Fig.~\ref{fig:xsample}
shows the estimates of $\beta$ (bottom left) and the evidence (bottom
right) over the 100 simulation runs after the $10^{\rm th}$
iteration. We find for the mean and 68\% confidence $\beta = 0.02998
\pm 0.00002$ and $\ln E = -264.0342 \pm 0.0010$. Comparing these
results to the posterior mean values corresponding to the sample of
$M=100$ simulated data points, $\beta = 0.0299862$ and $\ln E =
-264.0344$ suggests that PMC performs very well in providing an
accurate estimate of the evidence and of $\beta$.  If we use only
$N=1\,000$ sample points per iteration, we still get stable and
reliable results, but with a fourfold increase of the error bar of
$\ln E$.

\subsection{Known twist $\beta$}

For the second case, we use the slightly twisted centred multivariate
Gaussian as before but this time $\beta$ is known, $\beta
=0.03$ as before, and we integrate over the posterior represented by
(\ref{eqn:banana}), i.e.
\begin{equation}
  E=\int \pi(x|\beta,\Sigma) \, \dd x
\end{equation}
is our target.
This presents a much more difficult exercise with the tails of the
posterior being a significant challenge to capture, and the
dimension of the space to be integrated over is $d=5$. To
estimate the evidence with PMC in this second example, the number of
points at each iteration was $N=10000$, the number of PMC iterations
$T=10$ and as in the first example we use 100 simulation runs to
assess the variability of the estimates. After the 10$^{\rm th}$ iteration we
draw a final sample of size $N=100,000$. The initial importance function
$q^1$ is a mixture of multivariate Student-t distributions with
components displaced randomly in different directions slightly away
from the centre of the range for each variable: the mean of the
components is drawn from a p-multivariate Gaussian with mean 0 and
covariance equal to $\Sigma_0/5$ where $\Sigma_0$ is 
the covariance of the proposal components.
We choose a mixture of 9 components of
Student-t distribution with $\nu = 9$ degrees of freedom; and
$\Sigma_0$ is a diagonal matrix with diagonal entries
$(200,50,4,\dots,4)$. This choice of ($\nu$, $\Sigma_0$) ensures
adequate coverage, albeit somewhat overdispersed, of the feasible
parameter region.

\begin{figure}
  \resizebox{\hsize}{!}{
    \includegraphics[bb=10 20 546 494]{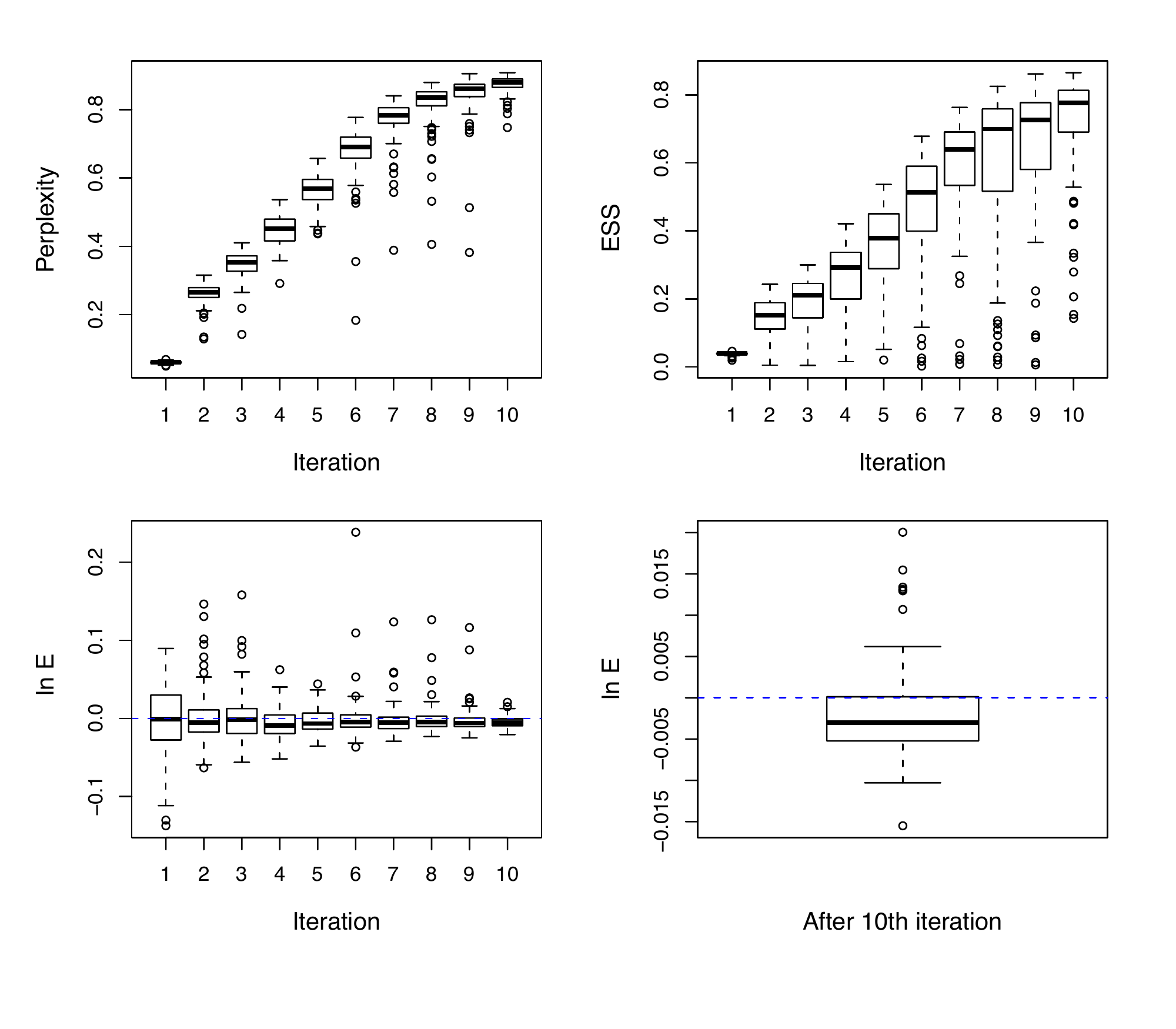}
  }
  \caption{PMC sampling of (\ref{eqn:banana}) as a function of
    iteration over 100 simulation runs. \emph{Top left:} Estimates of
    the perplexity (\ref{perplexity}); \emph{Top
      right:} Effective sample size (\ref{ess}); both normalized by $N$.
    \emph{Bottom:} Estimates of the evidence at each iteration
    (\emph{left}), and after the 10$^{\rm th}$ iteration using
    $N=100,000$ sample points (\emph{right}). The true value $0$ is
    indicated as a dashed line. See Fig.~\ref{fig:xsample} for
    details about the whisker plot representation.}
  \label{fig:res-target-evid}
\end{figure}

Fig.~\ref{fig:res-target-evid} shows the estimates of the evidence
(bottom right) against the true value over the 100 simulation runs
after the 10$^{\rm th}$ iteration.

The results for this more difficult case suggest that PMC performs
reasonably well in providing an accurate determination of the
evidence.  The estimates at each iteration
(Fig.~\ref{fig:res-target-evid}, bottom left) are stable, with a
reduction in the variability seen as the importance function better
adapts to the posterior density. The adaptation performance can
be seen by the increase in estimates of the normalized perplexity
and effective sample size (Fig.~\ref{fig:res-target-evid}, top right)
for successive iterations. See fig.~1 of WKB09 for a graphical example
of the adaptation of the importance function to the posterior for
$d=10$. After the final iteration (Fig.~\ref{fig:res-target-evid},
bottom right), the estimate of the evidence is $\ln E = -0.0019 \pm
0.0038$ at 68\% confidence. The presence of a slight downward bias
from the true value of zero is not unexpected due to the use of the
log-scale (as a consequence of Jensen's inequality, an unbiased
estimate of $E$ would look biased when plotted on the log-scale). This
could also be due to the importance function not fully exploring the
low probability tails of the posterior density. However in the case of
Fig.~\ref{fig:res-target-evid}, the bias in terms of the scale is very
small, and is associated with equally small variability of the
estimates of $E$.


\section{Cosmology}
\label{sec:cosmo}

The so-called standard model of cosmology is successful in explaining
recent observations of cosmology, such as the CMB, supernovae of type
Ia (SNIa), galaxy clustering including baryonic acoustic oscillations
(BAO), cosmic shear, galaxy cluster counts, and Ly$\alpha$ forest
clustering. This flat $\Lambda$CDM model has only six free parameters
$(\Omegam, \Omegab, h, n_{\rm s}, \tau, \sigma_8$, or functions
thereof) and is therefore surprisingly simple.

Despite this, various extensions to the standard model are
considered and tested routinely using observational data.  These
extensions may be based on independent evidence (for example massive
neutrinos from oscillation experiments), be predicted by a compelling
hypothesis (primordial gravitational waves from inflation) or reflect
our ignorance about the fundamental physics (dynamical dark energy).
Whatever be the case, future surveys and analyses are to answer the
question which of the many models is the one to best describe the
observations. So far, no extension of the standard model has been strongly
supported by the data.

In this paper, we use the Bayesian evidence as a tool to compare
different models and their ability to describe cosmological data. As
described in Sect.~\ref{sec:PMC}, we use PMC to sample the posterior
and to calculate the evidence. We take recent data of CMB
\citep{WMAP5-Hsinshaw08}, SNIa \citep{kowalski-2008} and BAO
\citep{2005ApJ...633..560E}. The extensions to the standard model of
cosmology concern dark energy and curvature
(Sect.~\ref{sec:de_models}), and inflationary
models (Sect.~\ref{sec:prim}).

\subsection{PMC set-up}
\label{sec:PMC-set-up}

To set up PMC we have to choose the initial proposal, $q^1$, the
number of sample points, $N$, and the number of iterations, $T$. We
take $q^1$ to be a Gaussian mixture model with $D$ components, which
are displaced from the maximum-likelihood point by a random shift
$f_{\rm shift}$ in each dimension, with $f_{\rm shift}$ being the
  fraction of the prior parameter range. The covariance
of the components corresponds to the Fisher matrix rescaled by a
number $f_{\rm var}$, typically of order unity, or larger if the
Fisher matrix is suspected to be significantly narrower than the
posterior curvature.

For the dark-energy and curvature models (Sect.~\ref{sec:de_models}),
we choose the number of iterations $T$ to be 10. If after 10
iterations the perplexity is still low, say, smaller than 0.6, we run
PMC for more iterations. The choices of the $N$ and $D$ are linked:
the average number of points sampled under an individual
mixture-component, $N/D$, should not be too small, to ensure a
numerically stable updating of this component. We choose $N=7\,500$
and $D=10$.

For the primordial models (Sect.~\ref{sec:prim}) we take $T=5,
N=10\,000$ and $D$ between 7 and 10, depending on the dimensionality
and shape of the likelihood.

The parameters controlling the initial mixture means and covariances,
are chosen for both cases to be $f_{\rm shift} = 0.02$, and $f_{\rm
  var}$ between 1 and 1.5. Those values are educated guesses and can
be refined according to the evolution of the proposal components
during the first few iterations. E.g., if many components die off
early, they start too far from the high-posterior region and $f_{\rm
  shift}$ should be decreased. For the final iteration we choose a
five-times larger sample than for previous iterations.

\subsection{Dark energy and curvature}
\label{sec:de_models}

Here we test the standard $\Lambda$CDM-model assumptions of a
cosmological constant and flatness. We parametrize the dark energy
equation-of-state parameter as constant and as linear function in the
scale factor $a$, respectively. Together with the basic model for which $w=-1$,
we compare the three cases:
\begin{align}
  w & = -1 & \Lambda{\rm CDM} \nonumber \\
  w & = w_0 & w{\rm CDM} \\
  w & = w_0 + w_1 (1-a) & w(z){\rm CDM} \nonumber
\end{align}
In addition, the curvature parameter $\Omega_K$ for each of the above
models is either $\Omega_K = 0$ (`flat') or $\Omega_K \ne 0$
(`curved').

We do not take into account dark energy clustering.
The observational
data is reduced to purely geometrical probes of the Universe; for CMB
these are the distance priors \citep{WMAP5-Komatsu08} and for BAO the
distance parameter $A$ \citep{2005ApJ...633..560E}.  The common
parameters for all models are $\Omegam, \Omegab$ and $h$. All models
share the same flat priors for those three parameters; the prior
ranges for all parameters can be found in Table
\ref{tab:priors_de_curvature}.
We verified that the relative evidence between models, or Bayes
factor, does not depend on the prior ranges for nested parameters if the
high-density likelihood region is situated far from the prior
boundaries.

\begin{table}
  \caption{Prior ranges for dark energy and curvature models. In case
    of $w(a)$ models, the prior on $w_1$ depends on $w_0$, see
    Sect.~\ref{sec:deprior}.}
  \label{tab:priors_de_curvature}
  \begin{tabular}{l|l|l|l}
    Parameter & Description & Min. & Max. \\ \hline
    $\Omegam$ & Total matter density & 0.15    & 0.45    \\
    $\Omegab$ & Baryon density & 0.01    & 0.08    \\
    $h$       & Hubble parameter & 0.5      & 0.9   \\ \hline
    $\Omega_K$ & Curvature  & $-1$   & 1     \\
    $w_0$      & Constant dark-energy par. & $-1$   & $-1/3$    \\
    $w_1$      & Linear dark-energy par. & $-1-w_0$ & $\frac{-1/3-w_0}{1-a_{\rm acc}}$   \\
  \end{tabular}
\end{table}

\subsubsection{Dark-energy prior}
\label{sec:deprior}

The simple parametrization of $w$ clearly is not motivated by
fundamental physics of dark energy. However, this choice represents
the most simple models which go beyond a cosmological constant; it
therefore makes sense to use those extensions in a model-selection
framework. To define a physically sound prior of these dark-energy
parameters, we restrict ourselves to a specific class of models. Our
goal is to find a model which is able to explain the observed, recent
accelerated expansion of the Universe. The model should therefore
include a component to the matter-density tensor with $w(a)<-1/3$ for
values of the scale factor $a>a_{\rm acc}$. We choose $a_{\rm
  acc} = 2/3$. To limit the equation of state from below, we impose
the condition $w(a) > -1$ for all $a$, thereby excluding phantom energy
as in \cite{2008MNRAS.388.1314E}. Fig.~\ref{fig:w0_w1} shows the allowed range
in the case of two dark-energy parameters.  We note that our approach
is inconsistent to some extend in that the data on which the
observation of accelerated expansion is based on is part of the data
used in this analysis.

\subsubsection{Curvature prior}

A natural limit on the curvature is that of an empty universe; this
certainly places an upper boundary on the curvature, corresponding to
$\Omega_K = 1$.  A lower boundary, corresponding to an upper limit on
the total matter-energy density, is less stringent. We choose
$\Omega_K>-1$; this `astronomer's prior' \citep{2009MNRAS.397..431V}
provides a symmetric prior around the null-hypothesis value and
excludes high-density universes which are ruled out by the age of the
oldest observed objects.

An alternative prior on $\Omega_K$ could be derived from the paradigm
of inflation.  However, most inflationary scenarios imply the
curvature to be extremely close to zero, on the order of
$10^{-60}$. The likelihood over such a prior on $\Omega_K$ is
essentially flat for any current and future
\citep{2008arXiv0804.1771W} experiment.  A model with such an
uninformative likelihood would be indistinguishable in terms of the
Bayesian evidence with respect to a flat model.

\subsubsection{Results}

\begin{figure}

  \resizebox{\hsize}{!}{
    \includegraphics[bb=70 0 290 230]{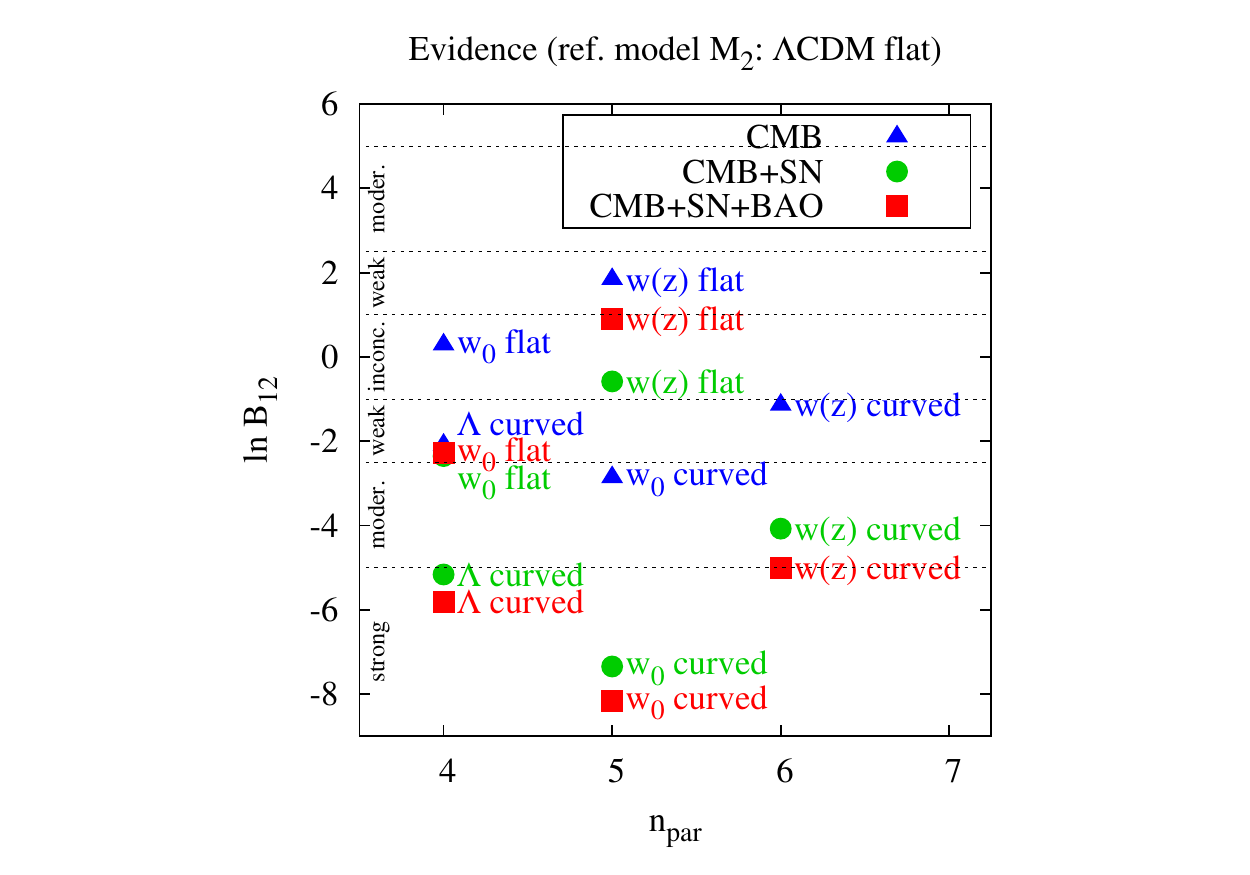}
  }

  \caption{Evidence for various models $\mathfrak{M}_1$ with respect
    to the reference flat $\Lambda$CDM model $\mathfrak M_2$ with
    $n_{\rm par} = 3$.  The different combinations of data are CMB
    (blue triangles), CMB+SNIa (green circles) and CMB+SNIa+BAO (red
    squares). Note that the Bayes factor between different
    non-reference models can only be compared for the same combination
    of data (same symbols).  }
  \label{fig:evi_w_curvature}
\end{figure}

In Fig.~\ref{fig:evi_w_curvature} we plot the Bayes factor for various
models with respect to the standard flat $\Lambda$CDM model.
In most cases there is positive evidence in favour of the standard
model. This evidence against more complex models increases if more
probes are combined. This is not surprising: no deviation from $w=-1$
and $\Omega_K=0$ has been found, additional parameters are not
supported by the data. The more data are added, the tighter get the
constraints around the standard values, therefore the stronger gets
the evidence in favour of this simplest model.

\begin{figure}

  \begin{center}
    \resizebox{0.95\hsize}{!}{
      \includegraphics{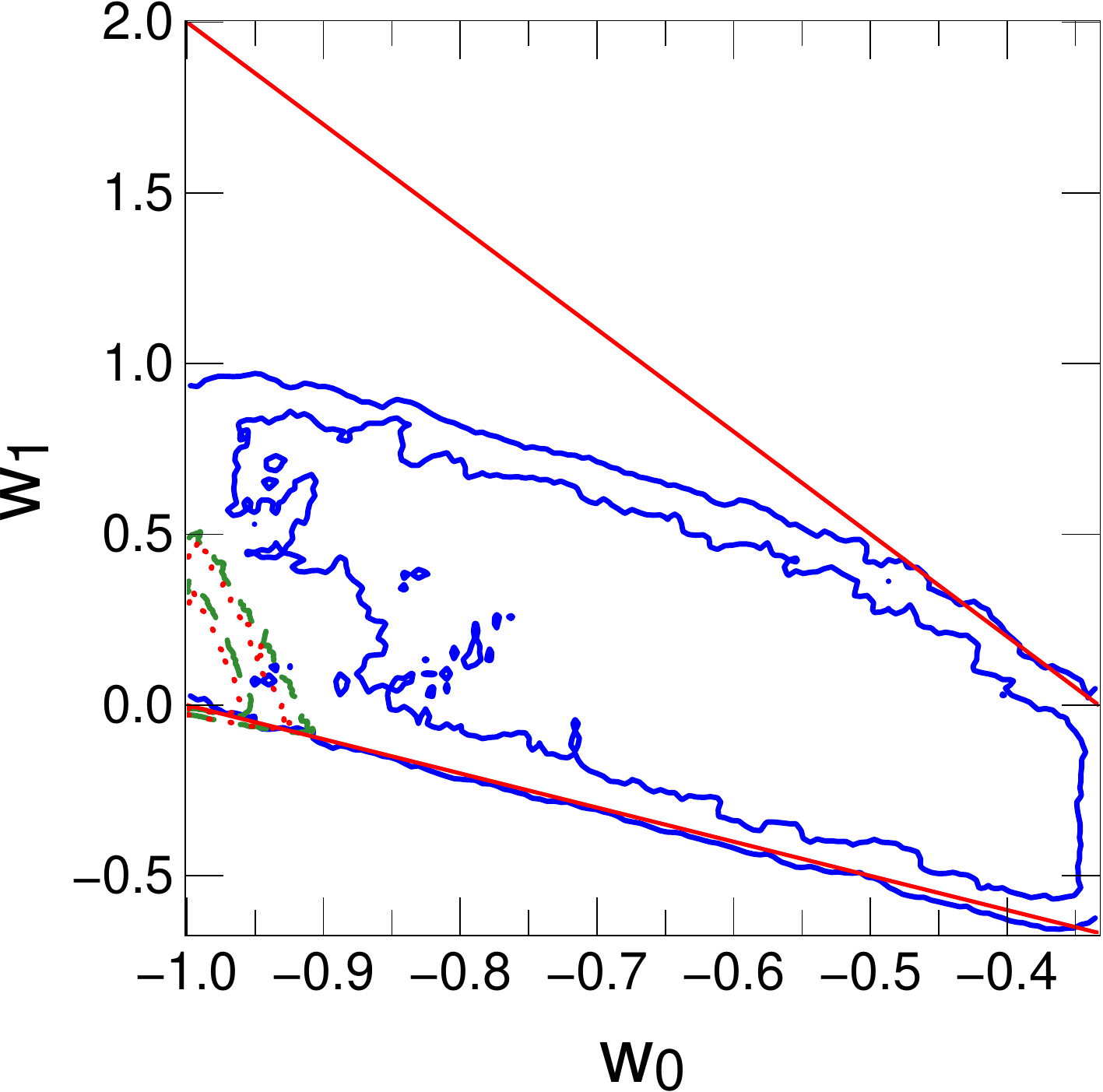}
    }
  \end{center}

  \caption{68\%- and 95\% confidence regions for WMAP (solid blue
    lines), WMAP+SNIa (dashed green) and WMAP+SNIa+BAO (dotted red curves).
    The allowed range for the dark-energy parameters $w_0$ and $w_1$
    lies between the two red straight lines.}
  \label{fig:w0_w1}
\end{figure}

The largest positive evidence is $\ln B_{12} = 1.8$, for the $w(z)$CDM
model and CMB alone. In this case, as can be seen in
Fig.~\ref{fig:w0_w1}, a large part of the prior range is still allowed
by the data, and a region of comparable size is excluded. There is
weak evidence that the two extra-parameters $w_0$ and $w_1$ are indeed
required by the data.  When adding SNIa and BAO, most of the prior
range is excluded, and this `waste' of parameter space is penalised by
decreasing the Bayes factor.

Regarding the prior on $w_0$, our $w$CDM model corresponds to Model II
from \cite{2007MNRAS.379..169S}. In that work, SN data alone led to a
$\ln B_{12}$ of around $-0.2$ (comparing $\Lambda$CDM to Model
II). In our case, the combination of SN with CMB and BAO leads to a
larger evidence in favour of $\Lambda$CDM.

Our results on the curvature are compatible with the findings of
\cite{2009MNRAS.397..431V}. Using all three data sets, a non-flat
universe is strongly disfavoured for all three dark-energy cases.  For
comparison, \citet{2009MNRAS.397..431V} showed that using a flat prior
in $\log \Omega_K$, corresponding to a flat prior on the curvature
\emph{scale}, therefore largely increasing the prior volume, leads to
inconclusive evidence.

\subsubsection{Stability of the results}

We test the reliability of the results for two of the cases
presented in Sect.~\ref{sec:de_models}, $w$CDM flat and $w$CDM curved,
both using CMB+SN+BAO data. We repeat the PMC runs 25 times. For
a given scheme with fixed proposal parameters $f_{\rm shift}, f_{\rm
  var}$ and $D$ (Sect.~\ref{sec:PMC-set-up}), we randomly vary the
positions and widths of the initial proposal components.

\begin{figure}
  \center
  \begin{tabular}{cc}
    {\includegraphics[bb=25 40 350 330, scale=0.35]{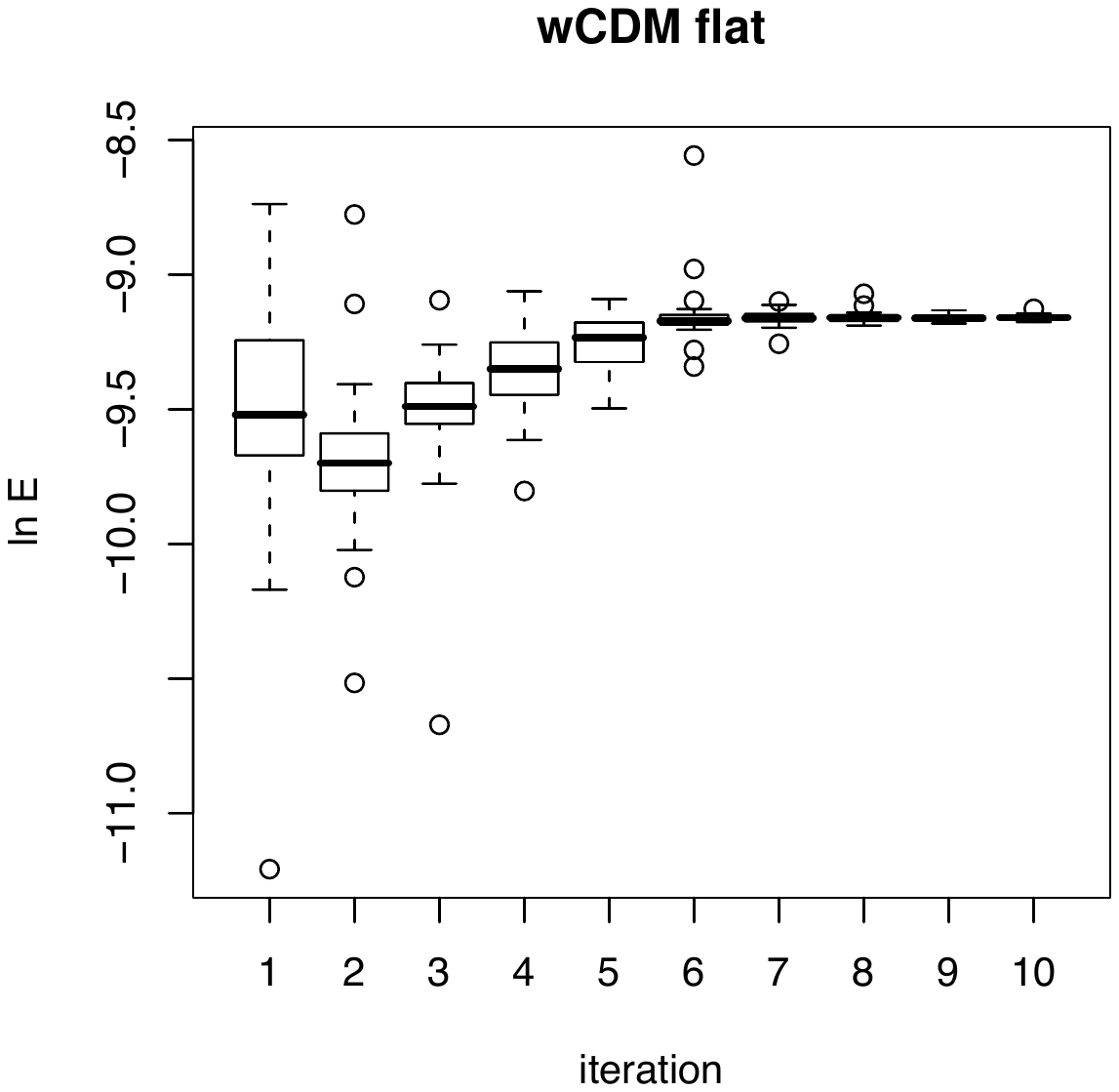}} &
    {\includegraphics[bb=25  40 300 330, scale=0.35]{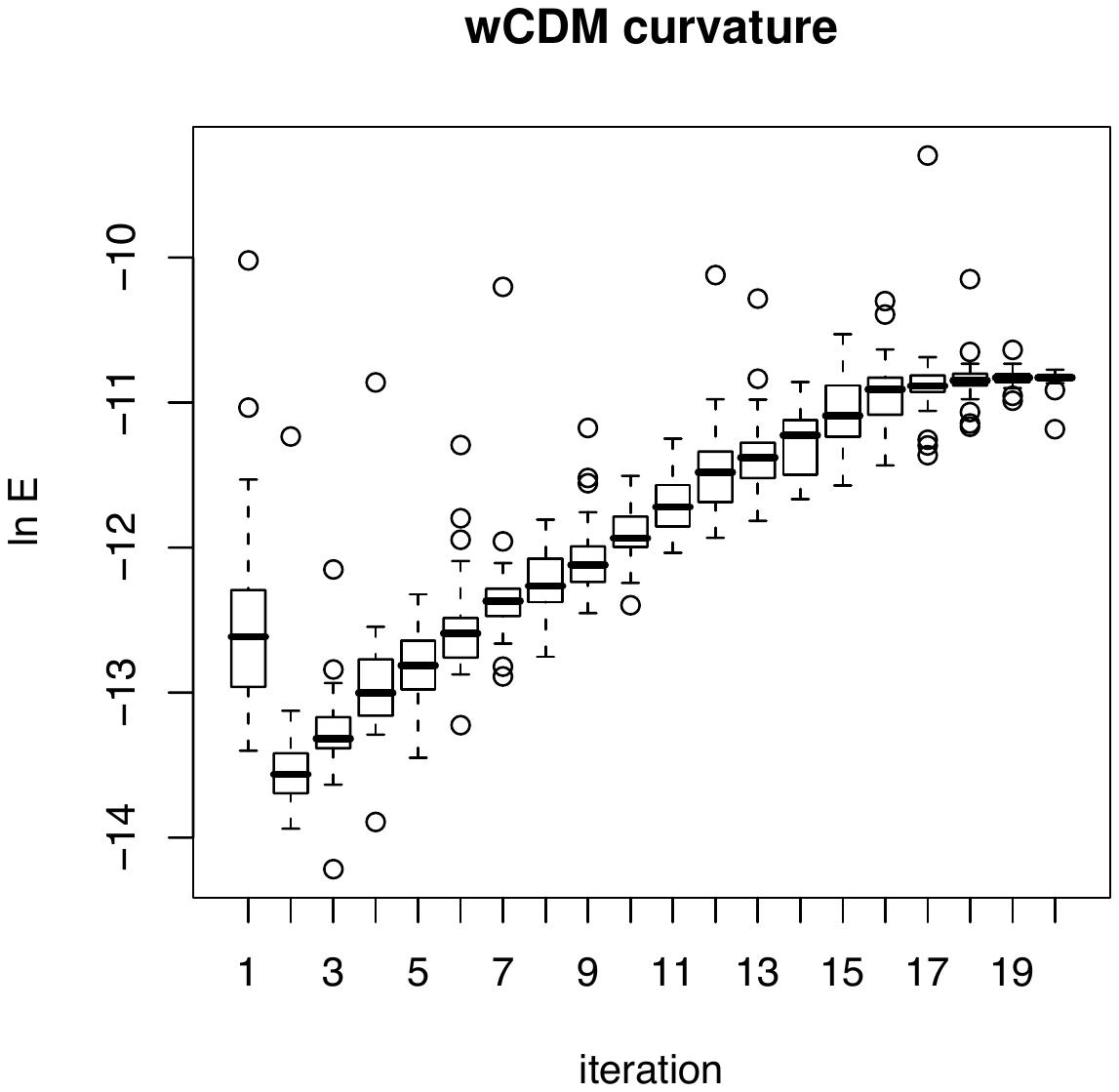}}\\[7ex]
    {\includegraphics[bb=25  40 350 330, scale=0.35]{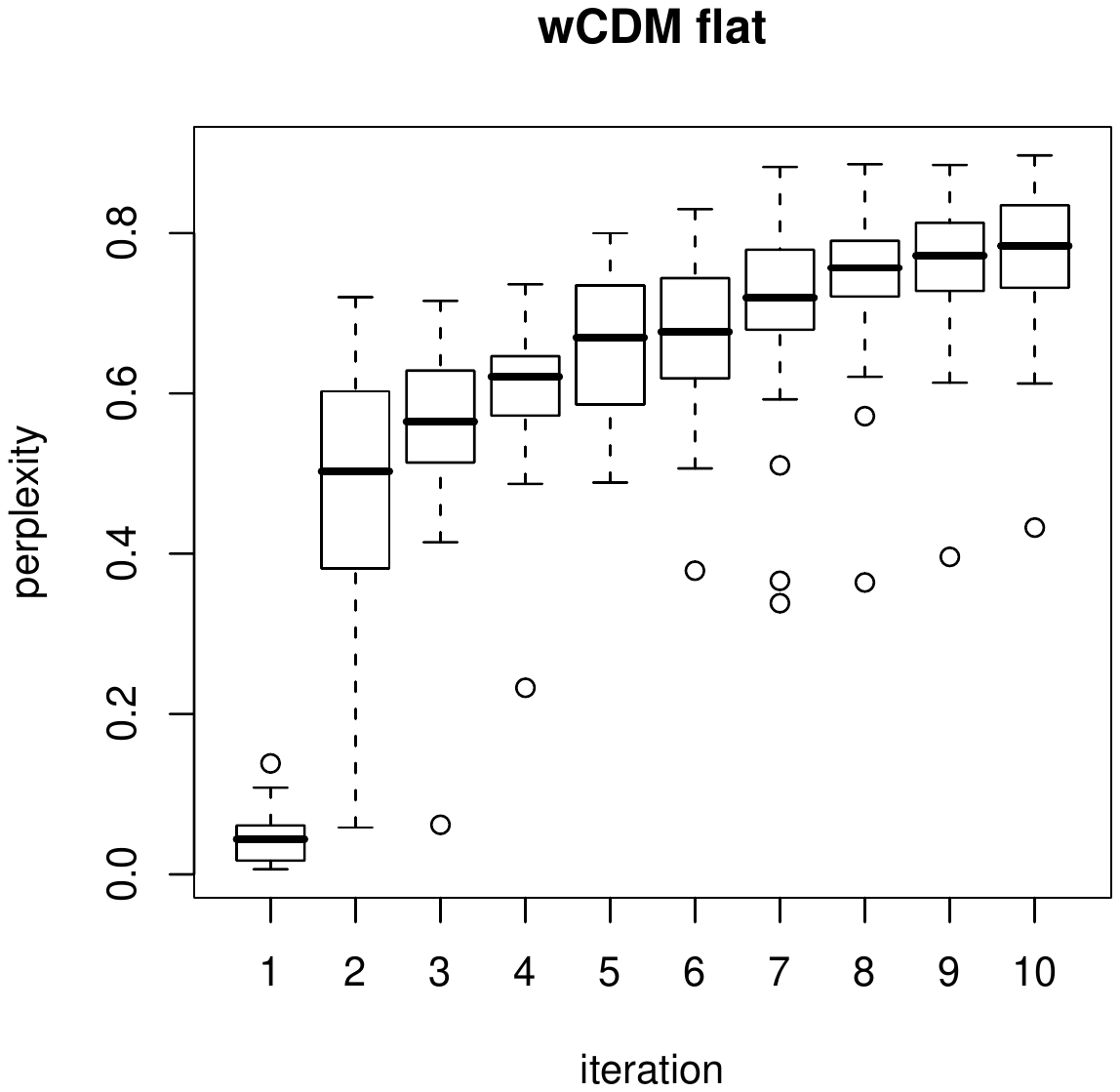}} &
    {\includegraphics[bb=25 40 300 330, scale=0.35]{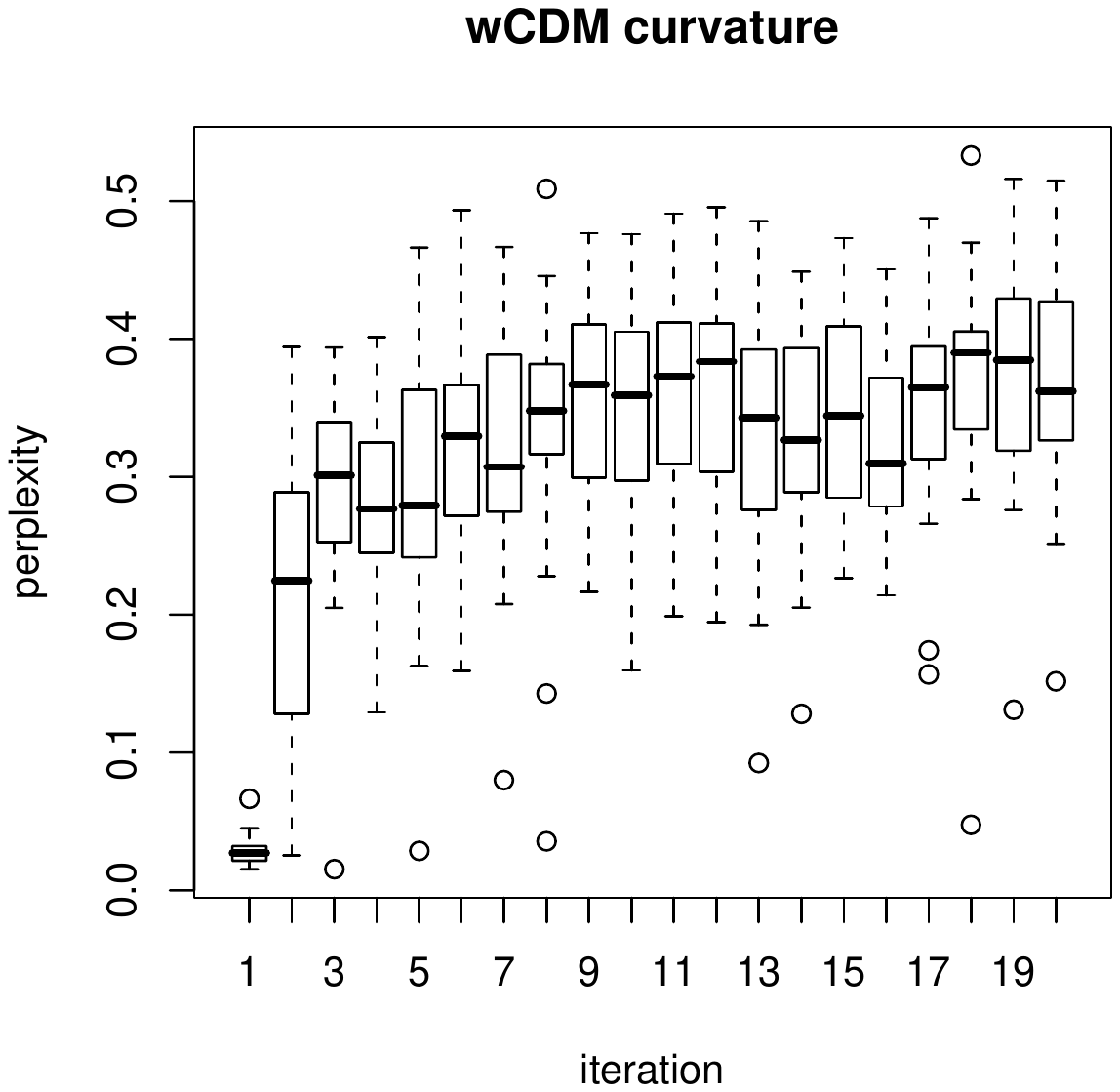}}\\[4ex]
  \end{tabular}
  \caption{Distribution of 25 PMC samplings of two dark-energy models,
    flat $w$CDM (\emph{left panels}) and curved $w$CDM (\emph{right
      panels}). \emph{Top (Bottom)}: The log-evidence (perplexity) as
    function of iteration. See Fig.~\ref{fig:xsample} for
    details about the whisker plot representation.}
  \label{fig:evidence.matrix}
\end{figure}

The distribution of the log-evidence $\ln E$ and the normalized perplexity is
shown in Fig.~\ref{fig:evidence.matrix} for two cases of dark-energy
models. Since the components of the proposal move towards the tails of
the posterior with progressing iteration, the evidence keeps on
increasing with better sampling of the tails.  The high value at the
first iteration is biased and dominated by a few points with very
large importance weights $w_n = \pi(x_n)/q(x_n)$, which are sampled
from the proposal tails but lie in regions of high posterior density
$\bar{\pi}$.

For the flat $w$CDM model the evidence converges after a few
iterations showing a very small dispersion between runs with $\ln E =
-9.159 \pm 0.011$ (68\% confidence) at $t=10$. The relatively large
perplexity of $p \approx 0.6$ -- $0.7$ indicates a reliable sampling
of the posterior. The posterior of the curved model is more elongated
which makes the sampling more difficult. The perplexity does not
exceed 0.4 even after 20 iterations. The evidence however stabilises
onto a narrow interval after $\sim 18$ iterations, $\ln E = -10.84 \pm
0.077$ (68\% confidence) at $t=20$.

\subsection{Primordial perturbations}
\label{sec:prim}

In the following, we test models corresponding to various descriptions
and parametrizations of primordial fluctuations. The (dark-matter)
density fluctuations are given by the power spectrum as function of
scale $k$,
\begin{equation}
  P_\delta(k) \propto k^{n_{\rm s} + \frac 1 2 \, \alpha_{\rm s} \ln (k/k_0)},
\end{equation}
with the parameters $n_{\rm s}$ being the scalar spectral index, and
$\alpha_{\rm s}$ the `running' of the index, i.e.~the
first-order Taylor term of the exponent. The pivot scale $k_0$ is
fixed to $k_* = 0.002$
Mpc$^{-1}$. In addition, tensor-modes (gravitational waves) have the power
spectrum
\begin{equation}
  P_{\rm t}(k) \propto k^{n_{\rm t}},
\end{equation}
with tensor spectral index $n_{\rm t}$. The ratio between tensor and
scalar perturbation spectra at scale $k_0$ is denoted by $r$. In the
standard model, $\alpha_s = n_{\rm t} = r = 0$, only $n_{\rm s}$ is a
free parameter.  Most inflationary models predict $n_{\rm s}$ to be
slightly below unity, therefore the power spectrum of primordial
density perturbations is a near scale-free power law.  Tensor
perturbations (gravitational waves) are expected to be non-zero, but
their amplitude is unknown and current data have not been able to
detect those modes.

Although tensor-modes are expected from most models of the early
Universe, they are not detected so far with the given sensitivity of
current data.

The models we consider for our Bayesian evidence analysis are
interpreted within the slow-roll approximation of inflation, as will
be described in the next section.

\subsubsection{Slow-roll parameters}

The slow-roll approximation of inflation provides an infinite
hierarchy of flow equations describing the dynamics of the single
scalar field which drives inflation \cite[see][and references
  therein]{2006JCAP...07..002P}. The slow-roll parameters $\epsilon$
and $^\ell\lambda_H, \ell \le 1$ are defined in terms of the potential $V$ of the
scalar field $\phi$, and the Hubble parameter $H$,
\begin{align}
  \epsilon & = \frac{m_{\rm Pl}^2}{4\pi} \left[ \frac {H^\prime} H \right]^2; \\
  ^\ell\lambda_H & = \left(\frac{m_{\rm Pl}^2}{4\pi}\right)^\ell
  \frac{\left(H^\prime\right)^{\ell-1}}{H^\ell} \frac{\dd^{\ell+1}
    H}{\dd \phi^{\ell+1}} ; \ell \le 1,
  \label{slow-roll-params}
\end{align}
where the prime denotes derivation with respect to $\phi$. The Planck
mass is denoted by $m_{\rm Pl}$.
The hierarchy of flow parameters can be truncated, since if some
$^L\lambda_H = 0$, all higher terms $^\ell\lambda_H, \ell>L$ vanish.
We consider the expansion up to first-order and set $^1\lambda_H =
\eta$, $^2\lambda_H = 0$. The parameters of the primordial power
spectra can be written in terms of the slow-roll parameters as
\begin{align}
  n_{\rm s} & = 1 + 2\eta - 4\epsilon - 2(1+{\cal C}) \epsilon^2 -
  \frac 1 2 (3 - 5 {\cal C}) \epsilon \eta;
  \label{ns_sr} \\
  r        & = 16 \epsilon \left[ 1 + 2 {\cal C}(\epsilon -
    \eta)\right];
  \label{r_sr} \\
  \alpha_{\rm s}  & = \frac \epsilon {1-\epsilon} \left( 10
    \eta - 8 \epsilon \right);
    \label{alpha_sr} \\
    n_{\rm t} & = -2 \epsilon - (3 + {\cal C}) \epsilon^2 + (1 + {\cal
      C}) \epsilon \eta.
    \label{nt_sr}
\end{align}
Here, ${\cal C} = 4 (\ln 2 + \gamma) - 5 \approx 
0.0814514$ where $\gamma = 0.577216$ is the Euler-Mascheroni
constant. For slow-roll inflation to take place, the slow-roll
conditions $\epsilon \ll 1$ and $|^\ell \lambda_H| \ll 1$ for all
$\ell$ have to be satisfied.

\subsubsection{Priors}

We use the slow-roll conditions to define priors on the primordial
parameters as $0 \le \varepsilon \le 0.1$ and $|\eta| \le 0.1$. Although
the exact values of the prior boundaries are somewhat arbitrary, they
have been considered by \citet{2006JCAP...08..009M} as natural limits
for the validity of the Taylor-expansion of the power spectrum $P(k)$
in $\ln(k/k_*)$ around the pivot scale $k_* = 0.05$ Mpc$^{-1}$. We use
$k_* = 0.002$ Mpc$^{-1}$ as pivot, in accordance to the WMAP5
analysis; we verified the equivalence of our results for both cases of
$k_*$.  We choose an uninformative (i.e.~flat) prior on the slow-roll
parameters. From eqs.~(\ref{ns_sr}-\ref{nt_sr}) we get the
corresponding ranges of the power-spectra parameters, see Table
\ref{tab:priors_primordial}, which are now motivated from fundamental
physical principles within the slow-roll model of inflation. We choose
flat priors for the power-spectra parameters as well, although they
are non-linearly related to the slow-roll parameters and the prior
will have a different shape. However, we ignore this for simplicity.
See \citet{2007MNRAS.378...72T} for a similar approach to define a
prior on the spectral index tilt. The tensor index $n_{\rm t}$ is
unconstrained by current data; therefore, we do not include this
parameter.

\begin{table}
  \caption{Prior ranges for primordial model comparison. The prior
    ranges for
    primordial parameters are derived from the slow-roll approximation.}
  \label{tab:priors_primordial}
  \begin{tabular}{l|l|l|l}
    Parameter & Description & Min. & Max. \\ \hline
    $\Omegam$ & Total matter density & 0.01    & 0.6    \\
    $\Omegab$ & Baryon density & 0.01    & 0.1    \\
    $\tau$    & Reionisation optical depth & 0.01    & 0.3    \\
    $10^9 \Delta_{\cal R}^2$ & Normalisation & 1.4 & 3.5 \\
    $h$       & Hubble parameter & 0.2      & 1.4   \\ \hline
    $n_{\rm s}$  & Scalar spectral index & 0.39   & 1.2 \\
    $\alpha_{\rm s}$ & Running of spectral index & -0.2 & 0.033 \\
    $r$ (lin.~prior) & Tensor-to-scalar ratio & 0 & 1.65 \\
    $\ln r$ (log.~prior)   & Tensor-to-scalar ratio & -80 & 0.50 \\
 \end{tabular}
\end{table}

We compare various models to the standard paradigm, which now has six
parameters ($\Omegam, \Omegab, h, n_{\rm s}, 10^9 \Delta_{\cal R}^2,
\tau)$.
For our model testing, we
single out individual parameters or combinations thereof. A strictly
consistent and thorough treatment should treat the slow-roll parameters
as primary parameters; this will be left for a future analysis.

\subsubsection{Results}

\begin{figure}
  
  \resizebox{\hsize}{!}{
   \includegraphics[bb=65 0 300 250]{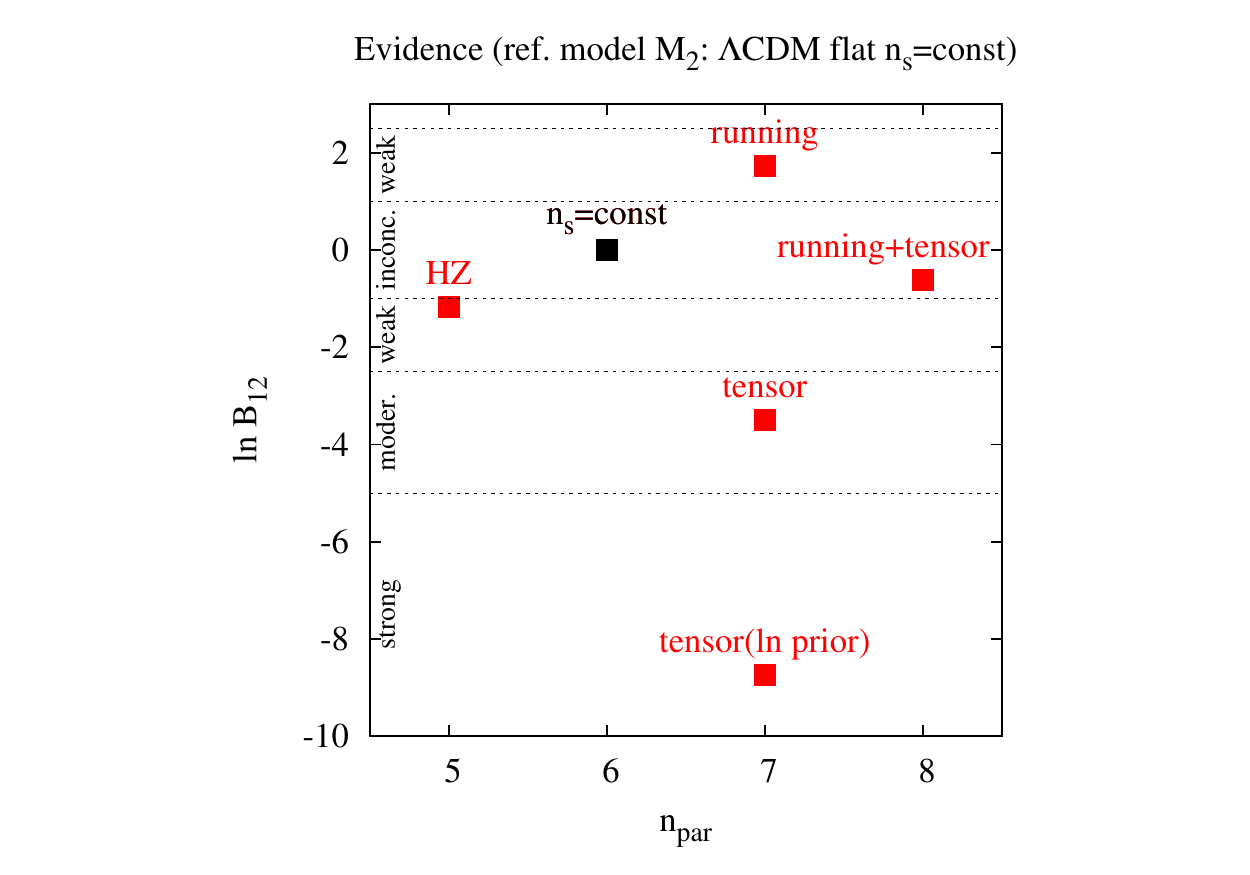}
  }
  
  \caption{Evidence for various models $\mathfrak{M}_1$ with respect to the
    reference model $\mathfrak{M}_2$, a flat $\Lambda$CDM universe with constant
    $n_{\rm s}$.}
  \label{fig:evi_primordial}
\end{figure}

In Fig.~\ref{fig:evi_primordial} we show the Bayes factor of various
models $\mathfrak{M}_1$ with respect to the standard model
$\mathfrak{M}_2$, a flat $\Lambda$CDM universe with $n_{\rm
  s}=\mbox{const}$. A running spectral index is favoured weakly, all
other cases are disfavoured. The evidence against the
Harrison-Zel'dovich model ($n_{\rm s}=1$) is weak, whereas tensor
perturbations are moderately disfavoured. For illustration, we include
a tensor-mode model with flat prior for $\ln r$ instead of $r$; the
minimum is chosen to be -80, corresponding to the energy scale
  of Big Bang Nucleosynthesis as a conservative lower limit of the
  inflation energy scale \citep{2006PhRvD..73l3523P}. The large prior
of the logarithmic tensor-to-scalar ratio causes this model to be
strongly disfavoured. Note however that this example is not consistent
with flat priors for the slow-roll parameters; it rather corresponds
to a model in which very small slow-roll parameters are much more
likely than large ones.

Previous results showed both positive as well as negative evidence for
a scale-free, Harrison-Zel'dovich (HZ) spectrum compared to a tilted power
law
\citep{2006MNRAS.369.1123B,%
2006ApJ...638L..51M,%
2007MNRAS.381...68B,%
2007MNRAS.378...72T}.
The evidence in either direction is however moderate at most. More
narrow priors on $n_{\rm s}$ increase the evidence for the HZ
model. Our result is in agreement with those previous works, taken into
account the differences in the employed data and priors.  This shows
that the evidence against a scale-free spectrum is not yet substantial with
the current data, and it calls for physically motivated prior density.

Previous results showed positive and negative evidence for a non-zero
running of the spectral index $\alpha_{\rm s}$, depending on the prior
width \citep{2007MNRAS.381...68B}. Our positive value of $+1.73$ is
relatively high but still corresponds to only weak evidence in favour
of $\alpha_{\rm s}\ne 0$.

\section{Conclusion}
\label{sec:discussion}


The Bayesian evidence $E$ provides a mathematically consistent and
intuitive tool to compare different models and to choose between
  competing models. Its calculation in high-dimensional parameter
spaces is in general numerically challenging, as the posterior
distribution may be multimodal
  and/or show strong non-linear parameter-dependencies.

Simplification methods such as the Laplace approximation are not
sufficient and cannot replace the full integration of the posterior
over the parameter space. Laplace assumes a multivariate
(i.e.~single-peaked) Gaussian likelihood and a prior which is much
wider than the likelihood. If either of these assumptions is
violated, the approximation can give values of $\ln E$ which are wrong
by several dex. Although in many cases the likelihood might be close
to Gaussian, or can be brought in such a form by parameter
transformations, this is not possible in general, {and often} there
exist hard, physical priors which cut off the likelihood.

In this paper, we use Population Monte Carlo (PMC) to estimate the
Bayesian evidence. PMC is a new, adaptive importance sampling method
which offers an efficient and reliable way to obtain the Bayesian
evidence for non-trivial likelihood shapes. An expression for the
variance of the evidence estimator is introduced. We will leave a
study of an estimator of this variance for future work. Here, we
repeat PMC runs to study the distribution of the evidence, and show
the robustness of the results. If the initial proposal is badly
chosen the estimate of $E$ can be significantly off; such cases can
however be identified by the in-built diagnostic tools of PMC,
e.g.~the perplexity and effective sample size. In such a case, the
perplexity would remain very low and never reach values close to its
maximum of unity. Additional indications of an unsuited initial
proposal is the vanishing of most mixture component. Those components
do not cover part of the high-density posterior region, and it is very
unlikely that the remaining components will well approximate the
posterior. An illustration of a well-behaving case can be seen in
Figure 8 of WK09, where the components spread out nicely until they
remain stationary after a few iterations.

Other methods to estimate the Bayesian evidence have been proposed,
e.g.~\textsc{vegas} \citep{lepage78,2007MNRAS.379..169S} which
necessitates that the likelihood function can be brought into a separable
form. Another computationally efficient since to some extent
parallelizable algorithm is \mbox{(multi-)nested} sampling
\citep{skilling:2007a,2008MNRAS.384..449F,2009MNRAS.398.1601F}.
While being a special case of importance sampling \citep{RW09},
nested sampling draws from the prior instead from an importance function 
approximating the posterior as in PMC. See \cite{2007ASPC..371..224C,MR09} and 
\cite{RW09} for an overview of various methods of Bayesian evidence estimation including
Markov chain Monte Carlo methods.


We have applied Bayesian model selection to two domains of cosmology, the accelerated expansion
of the Universe in the recent past and primordial fluctuations in the
early Universe. For the former, we analysed simple, parametrized models
of dark energy; the latter used the slow-roll approximation of inflation.
We employed recent cosmological data corresponding to CMB, SNIa
and BAO.


No dark-energy model is strongly or even only moderately favoured over
the standard $\Lambda$CDM paradigm. This is in spite of the rather
strong prior on dark-energy, i.e.~excluding phantom energy and
requiring an accelerating component in the recent past. More
general dark-energy models with larger parameter spaces will likely be
disfavoured with respect to $\Lambda$CDM. This is even true if
future experiments find deviations of $w$ from -1 unless the error
bars get extremely small (Lindley-Jeffrey paradox). This should
serve as a motivation to define a tight physical framework for dark-energy
models with stronger prior constraints on parameters.

We find strong evidence against a non-flat universe using the combined
CMB+SNIa+BAO data. This is true regardless of the chosen dark-energy
model. It holds for a prior belief that $|\Omega_K| \le 1.$

We use a natural limit on slow-roll inflation parameters to deduce
prior ranges for the primordial perturbation spectra parameters. The
preferred model contains a non-zero running spectral index
($\alpha_{\rm s}\ne 0$) and no tensor modes ($r=0$). The evidence is
however only weak. A scale-free, Harrison-Zel'dovich model is weakly
disfavoured. Tensor modes are moderately to strongly
disfavoured, depending on the prior shape on $r$. As a consequence, future
detections of tensor modes have to be done with very high significance,
to strongly disfavour a $r=0$ model.

\section*{Acknowledgments}

We would like to thank Jean-Francois Giovannelli, David Parkinson,
Roberto Trotta and Jean-Phillip Uzan for fruitful discussions. We
thank the referee, Martin Hendry, for useful comments and suggestions
which helped to improve the paper. We acknowledge the use of the
Legacy Archive for Microwave Background Data Analysis
(LAMBDA). Support for LAMBDA is provided by the NASA Office of Space
Science. We thank the Planck group at IAP and the \textsc{Terapix}
group for support and computational facilities. MK and DW are
supported by the CNRS ANR `ECOSSTAT', contract number
ANR-05-BLAN-0283-04 ANR ECOSSTAT. This project is partly supported by
the Chinese National Science Foundation Nos. 10878003 \& 10778725, 973
Program No. 2007CB 815402, Shanghai Science Foundations and Leading
Academic Discipline Project of Shanghai Normal University (DZL805).

\bsp

\label{lastpage}

\bibliographystyle{mn2e}
\bibliography{astro}

\end{document}